\documentclass[aps,prd,nofootinbib,showkeys,preprint,floatfix]{revtex4}

\usepackage[utf8]{inputenc}
\usepackage{amssymb}
\usepackage{amsmath}
\usepackage{amsfonts}
\usepackage{graphicx}
\usepackage{color}
\usepackage{xspace}
\usepackage{ulem}
\usepackage{lscape}
\usepackage{ wasysym }
\usepackage{floatrow}
\usepackage{subfig}

\usepackage{multirow,rotating}

\def\gsim{\raise0.3ex\hbox{$\;>$\kern-0.75em\raise-1.1ex\hbox{$\sim\;$}}}
\def\lsim{\raise0.3ex\hbox{$\;<$\kern-0.75em\raise-1.1ex\hbox{$\sim\;$}}}

\newcommand{\ba}[1]{\begin{eqnarray} \label{(#1)}}
\newcommand{\ea}{\end{eqnarray}}

\newfloatcommand{capbtabbox}{table}[][\FBwidth]

\newcommand{\AddrAHEP}{
  {\it AHEP Group, Instituto de F\'{\i}sica Corpuscular --
    CSIC/Universitat de Val{\`e}ncia \\
    Parque Cient\'{\i}fico, C/Catedr\'atico Jos\'e Beltr\'an, 2, 
    E-46980 Paterna, Spain}}

\newcommand{\AddrUFSM}{ 
Departamento de F\'isica y CCTVal\\
Universidad T\'ecnica Federico Santa Mar\'ia, 
Casilla 110-V, Valpara\'iso, Chile}

  \newcommand{\AddrULS}{
Departamento de F\' isica y Astronom\' ia, Facultad de Ciencias, 
Universidad de La Serena, \\
Avenida Cisternas 1200, La Serena, Chile }

\def\gsim{\raise0.3ex\hbox{$\;>$\kern-0.75em\raise-1.1ex\hbox{$\sim\;$}}}
\def\lsim{\raise0.3ex\hbox{$\;<$\kern-0.75em\raise-1.1ex\hbox{$\sim\;$}}}

\begin{document}

\preprint{IFIC/19-20}  

\title{Exotic coloured fermions and lepton number violation at the LHC}

\author{E. Carquin}\email{edson.carquin@usm.cl}\affiliation{\AddrUFSM}
\author{N. A. Neill}\email{nicolas.neill@gmail.com}\affiliation{\AddrUFSM}
\author{J.C. Helo} \email{jchelo@userena.cl}\affiliation{\AddrULS}
\author{M. Hirsch} \email{mahirsch@ific.uv.es}\affiliation{\AddrAHEP}



\begin{abstract}
Majorana neutrino mass models with a scale of lepton number violation
(LNV) of order TeV potentially lead to signals at the LHC. Here, we
consider an extension of the standard model with a coloured octet
fermion and a scalar leptoquark. This model generates neutrino masses
at 2-loop order. We make a detailed MonteCarlo study of the LNV signal
at the LHC in this model, including a simulation of standard model
backgrounds. Our forecast predicts that the LHC with 300/fb should be
able to probe this model up to colour octet fermion masses in the
range of (2.6-2.7) TeV, depending on the lepton flavour of the final
state.

\end{abstract}

\maketitle

\section{Introduction}

All Majorana neutrino mass models with a scale of lepton number
violation (LNV) of roughly $\Lambda_{LNV} \sim {\cal O}$(TeV) can lead
to lepton number violating signals at the LHC. The best-known example
is the left-right symmetric extension of the standard model
\cite{Pati:1974yy,Mohapatra:1974gc,Mohapatra:1980yp}. Here,
right-handed $W_R$ boson production can lead to the final states
$\ell^{\pm}\ell^{\pm}jj$ and $\ell^{\pm}\ell^{\mp}jj$ \cite{Keung:1983uu}.

There are, however, many other possible electro-weak scale extensions
of the standard model that potentially lead to LNV signals at the
LHC. In particular, the systematic analysis of the short-range
contributions to neutrinoless double beta decay \cite{Bonnet:2012kh}
has found a variety of such models, all of which can in principle
explain neutrino oscillation data. (For a recent global fit of all
oscillation data see, for example \cite{deSalas:2017kay}.)  Rough
estimates of the LHC reach, compared with the sensitivity of current
and future double beta decay experiments, have been made in
\cite{Helo:2013dla,Helo:2013ika,Gonzales:2016krw}. In this paper, we
study LHC signals for a particularly simple LNV extension of the
SM. This model generates neutrino masses at 2-loop order and, thus,
one expects the masses of the exotic particles of this model to be at
least partially within reach of the LHC. Different from previous
papers \cite{Helo:2013dla,Helo:2013ika,Gonzales:2016krw}, here we
perform a full detector simulation and background study, in
order to give more realistic estimates for future LHC sensitvities.

Only very few searches for LNV final states at the LHC exist so
far. CMS \cite{Khachatryan:2014dka} has searched for same-sign
dileptons plus jets in 8 TeV data. The results were interpreted as
lower limits on the mass of $W_R$ as function of right-handed neutrino
mass.  Lower limits on $m_{W_R}$ approaching 3 TeV have been derived,
for $m_N < m_{W_R}$ and assuming the gauge coupling of the
right-handed bosons to be equal to the standard model $SU(2)_L$
coupling. A small excess around $p^2_{\ell\ell jj}=2$ TeV was observed
(statistically with a local significance of 2.8 $\sigma$), but
discarded by the experimentalists as a signal, since the sample
consists dominantly of opposite sign lepton final states. (See,
however, the discussion in \cite{Anamiati:2016uxp}.) Unfortunately,
the recent update of this search by CMS \cite{Sirunyan:2018pom} with
$\sqrt{s}=13$ TeV data, could not reproduce this excess and now quotes
a lower limit of $m_{W_R} \ge 4.4$ TeV. A very similar search by ATLAS
\cite{Aaboud:2018spl} has also been published, providing lower limits
extending up to $m_{W_R} \ge 4.7$ TeV, for $m_N \lsim m_{W_R}/2$.

We also want to mention that for models with a seesaw type-II, pair
production of the doubly charged component of the triplet can lead to
$\Delta^{++}\Delta^{--}\to \ell^{\pm} \ell^{\pm} W^{\mp}W^{\mp} \to
\ell^{\pm}\ell^{\pm}+4j$
\cite{Azuelos:2004mwa,Perez:2008ha,Melfo:2011nx}.\footnote{To 
establish LNV experimentally one needs to study final states
{\em without} missing energy.} If only leptonic or $WW$ final states
were observed, LNV could not be established at the LHC, but the type of
scalar multiplet could still be determined \cite{delAguila:2013yaa}. 
However, no search for $\ell^{\pm}\ell^{\pm}W^{\mp}W^{\mp}$ at the LHC
exists so far. Instead, ATLAS \cite{Aaboud:2017qph} searched for
$pp\to \Delta^{++}\Delta^{--} \to 4 \ell$. For a doubly charged Higgs
boson only coupling to left-handed leptons, the limits vary from (770 -
870) GeV, depending on the lepton flavour, assuming
Br($\Delta^{\pm\pm}\to \ell^{\pm}_{\alpha} \ell^{\pm}_{\beta}$) equal to 1
(for $\alpha,\beta=e,\mu$).  ATLAS \cite{Aaboud:2018qcu} has also
searched for $pp\to \Delta^{++}\Delta^{--} \to 4 W$. However, lower
limits, based on 36.1/fb, are currently only of order 220 GeV.

The model we consider in this paper contains two new particles: A
scalar leptoquark (LQ) and an exotic colour-octet fermion, $\Psi$ (for
details see section \ref{sec:model}). The fermion can be pair-produced,
decaying to the final state $\ell^{\pm}jj$. We will discuss
restrictions on the fermion in this model from the searches
\cite{Sirunyan:2018pom,Aaboud:2018spl} in section \ref{sec:results}.
As for the leptoquark, currently the best limits come from CMS
\cite{Sirunyan:2018btu} and ATLAS \cite{Aaboud:2019jcc}.  CMS
\cite{Sirunyan:2018btu} finds lower limits on pair-produced
leptoquarks, giving $m_{LQ}\ge 1435$ ($1270$) GeV for a branching
ratio of $\mbox{Br}(S_{LQ} \to \ell^{\pm}j)=1$ ($0.5$). ATLAS derives
\cite{Aaboud:2019jcc} a very similar number of $m_{LQ}\ge 1.25$ TeV
for a branching ratio equal to 0.5.
 
In order to estimate the reach of our model for the LHC Run-3, we
performed detector level studies of the same-sign dilepton plus four
jets final state. In section \ref{sec:mc} the
MonteCarlo simulation and the cut and count analysis optimization are
discussed in some detail. Signal and SM background $H_T$ distributions
are shown and the corresponding yield tables after the selection cuts
are included for completeness in the appendix (section \ref{sec:app}). 
In section \ref{sec:results}, our results are presented as the $2\sigma$
limits and $5\sigma$ discovery regions, forecasted for the
$\Psi\rightarrow \ell^{\pm}jj$ branching ratio, with $\ell=e,\mu$, as a
function of the colour octet fermion mass.

The rest of this paper is organized as follows: in section
\ref{sec:model} we describe the model basics, discuss briefly non-LHC
constraints, such as neutrino masses, and describe the benchmark
scenarios we use in the rest of the paper. In section \ref{sec:mc} the
MonteCarlo simulation is discussed and our results are presented in
section \ref{sec:results}.  In section \ref{sec:conclusions} the
conclusions and outlook of this work are given.

\section{Model Basics} \label{sec:model}

In this paper, we use a particularly simple 2-loop neutrino mass
model.  The model adds only two new particles to the standard model: A
(singlet) scalar leptoquark $S_{LQ} \equiv S_{3,1,-1/3}$ and a colour
octet fermion, $\Psi \equiv \Psi_{8,1,0}$.  Here the subscripts denote
the transformation properties/charge under the SM gauge group,
$SU(3)_c \times SU(2)_L \times U(1)_Y$.  We note that this model has
appeared twice in the literature before. It was listed in
\cite{Bonnet:2012kh}\footnote{Decomposition T-I-5-i for the Babu-Leung
operator \#11 \cite{Babu:2001ex}.} as one particular example of a
short-range contribution to neutrinoless double beta decay. And in
\cite{Angel:2013hla} predictions for neutrino masses and low energy
lepton flavour violation for this model have been worked out in
detail.

With these new fields the lagrangian of the model contains the 
following terms:
\begin{eqnarray}
{\cal L} &=& (Y_1)_{\alpha\beta} \overline{L^c}_\alpha Q_{\beta} S^\dag_{LQ} 
        + (Y_2)_{\alpha\beta} \overline{d_R}_\alpha \Psi_{\beta} S_{LQ} 
        + (Y_3)_{\alpha\beta} \overline{e_R^c}_\alpha u_{R,\beta} S^\dag_{LQ} 
        + h.c.
\\ \nonumber
&+& \frac{1}{2} m_\Psi \bar{\Psi}^c \Psi 
   + m_{LQ}^2 S^\dag_{LQ} S_{LQ} + \cdots
\end{eqnarray}
Here $\alpha$ and $\beta$ are generation indices and we have left open
the possibility that more than one copy of $\Psi$ could exist. The
quantum numbers of the scalar leptoquark $S_{LQ}$ allow, in principle,
to write down two more terms in the lagrangian, $ \bar{Q}Q^c S_{LQ}$
and $\bar{u} d^c S_{LQ}^\dag$. These, if coupled to the first
generation of quarks, induce rapid proton decay. As noted in
\cite{Baldes:2011mh}, stringent upper limits on Yukawa couplings in
any quark generation indices can be derived from the requirement of
successful baryogenesis.  In order to avoid these problems, we simply
postulate baryon-number conservation as an additional symmetry of the
model.

\begin{figure}[hbt]
\includegraphics[scale=0.9]{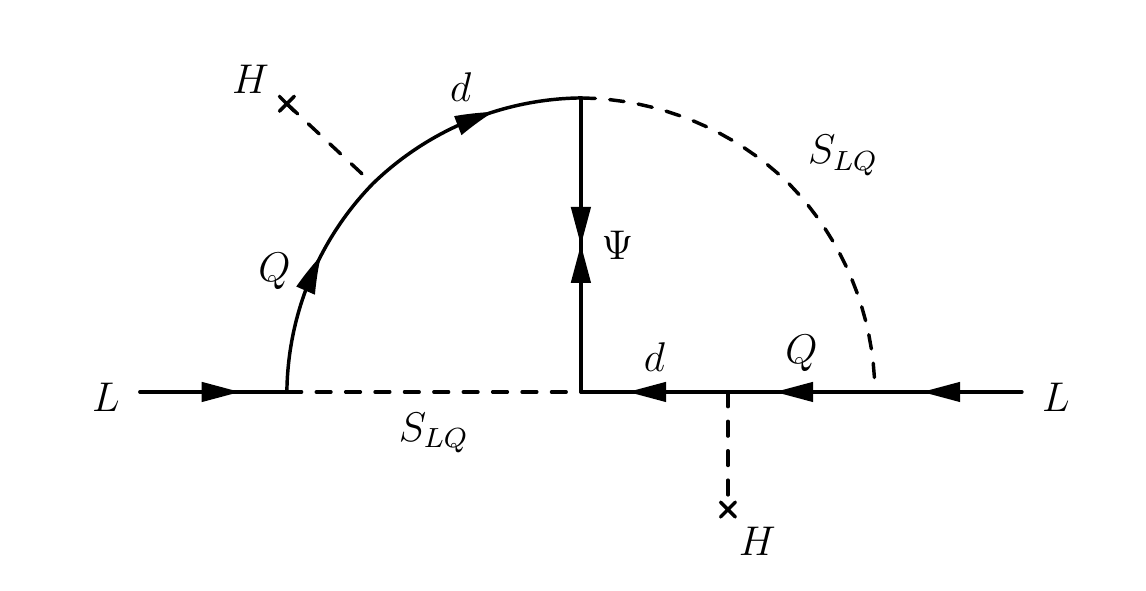}
\captionsetup{justification=raggedright,
singlelinecheck=false
}
\caption{2-loop neutrino mass diagram for the model considered in this
  work.\label{fig:diag}}
\end{figure}  

The main motivation for studying this model is that it can explain
neutrino oscillation data by generating neutrino masses at 2-loop
level as shown in figure (\ref{fig:diag}). Assuming only one copy of
$\Psi$ this diagram gives a contribution to the neutrino mass matrix,
roughly as \cite{Angel:2013hla,deGouvea:2007xp,Sierra:2014rxa}
\begin{eqnarray}
\label{eq:mnu}
(m_\nu)_{\alpha \beta} \sim \frac{N_c}{(16 \pi^2)^2 }
\frac{m_{\Psi}}{m_{LQ}^2} 
\Big[(Y_1)_{\alpha k} (Y_2)_{k} m_k I_{kr}(x^2) m_r (Y_2)_{r} (Y_1)_{\beta r}
+  \alpha \leftrightarrow \beta \Big].
\end{eqnarray}  
Here, $N_c$ is a color factor, $m_k$ and $m_r$ are down quark masses
of generation $k,r$ and $I_{kr}(x^2)$ stands for the 2-loop integral
\cite{Angel:2013hla,Sierra:2014rxa}, with the dimensionless argument
$x=(m_{\Psi}/m_{LQ})$. Note that with the quark masses much smaller
than the mass of $\Psi$ and $S_{LQ}$, $I_{kr}(x^2)$ is to a good 
approximation independent of $k$ and $r$, i.e. $I(x^2)$.  In order to
reproduce the neutrino mass, estimated from the atmospheric neutrino
masss scale ($m_{\nu} \sim 0.05$ eV), and assuming very roughly
$m_{\Psi} \sim m_{LQ} \sim \Lambda_{LNV} \sim {\cal O}$(TeV), the
Yukawa couplings (for third generation quarks) in eq. (\ref{eq:mnu})
should be of order (few) ${\cal O}(10^{-2})$.

Since neutrino mass models must not only produce the correct absolute
value of one neutrino mass, but also reproduce the solar mass scale
and the observed flavour structure of the neutrino mass matrix, the
above estimate is only indicative of the typical size of parameters.
On closer inspection, one sees that in the limit where only $m_b$ is
taken different from zero, eq. (\ref{eq:mnu}) generates only one
neutrino mass. The authors of \cite{Angel:2013hla} therefore suggested
to use two copies of $S_{LQ}$. However, we note that contributions
proportional to $m_s$, while being smaller than those proportional to
$m_b$, could generate a large enough second neutrino mass, if the
entries in the Yukawa matrices $Y_1$ and $Y_2$ for 2nd generation
quarks are roughly larger, by a factor $\sqrt{m_b/m_s} \sim 6$ each,
than those for 3rd generation quarks. Also, the fit of non-zero
neutrino angles requires flavor off-diagonal entries in
$(m_\nu)_{\alpha \beta}$, implying lepton flavour violating charged
lepton decays. We will not repeat this discussion here, since a
detailed study can be found in \cite{Angel:2013hla}.

\begin{figure}[t]
\includegraphics[scale=0.6]{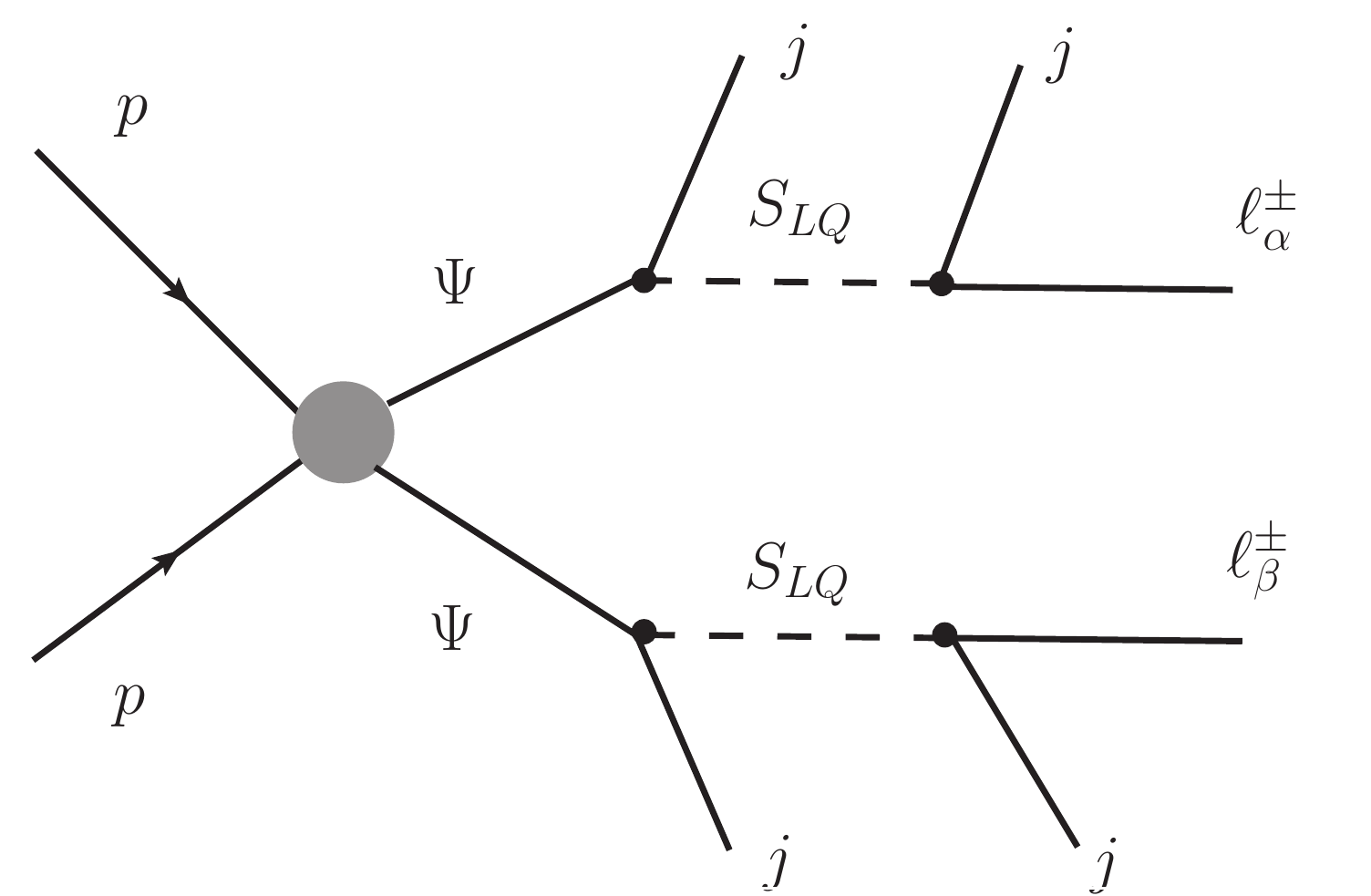}
\captionsetup{justification=raggedright,
singlelinecheck=false
}
\caption{Pair production of the colour-octet fermion $\Psi$, followed
  by 3-body decays, at the LHC.\label{fig:diagLHC}}
\end{figure}

Lepton number violation in this model is due to the Majorana mass term
$m_{\Psi}$. $\Psi$, once produced, can decay to a down quark and
$S_{LQ}$. Since $\Psi$ is a Majorana fermion, decays to
$\overline{d_R} S_{LQ}$ and $d_R S_{LQ}^{\dagger}$ are equally
likely. Thus, pair-produced $\Psi$ will lead to the LNV and LNC final
states $\ell^{\pm}_{\alpha}\ell^{\pm}_{\beta}+4j$ and
$\ell^{\pm}_{\alpha}\ell^{\mp}_{\beta}+4j$, as shown in
figure (\ref{fig:diagLHC}). There are, however, some important differences
to the case of the type-II seesaw discussed in the
introduction. First, $\Psi$ being a colour octet, production cross
sections are much larger in the current model. And, second, in type-II
seesaw the invariant masses of the subsystems
$\ell^{\pm}_{\alpha}\ell^{\pm}_{\beta}$ and $4j$ should both equal the
mass of $\Delta^{++}$. Here, on the other hand, the invariant masses
of two particular $\ell jj$ subsets should produce mass peaks. As discussed
in \cite{Helo:2013ika}, if a discovery of LNV is eventually made at the LHC,
this can be used to distinguish different models.

More important for our forecasts is, however, that the information
from neutrino oscillation experiments is not sufficient to fix all
entries in the Yukawa matrices $Y_{1,2,3}$. The large mixing angles
observed indicate that lepton flavour violating final states should
likely be large in $\ell^{\pm}_{\alpha}\ell^{\pm}_{\beta}+4j$. On the other
hand, the sensitivity of the LHC to final states involving tau leptons is
markedly less than for muons or electrons. We thus decided to consider
in our numerical studies only three simple benchmark scenarios for the
Yukawa couplings. These are:
\begin{enumerate}
  \item The scalar leptoquark couples to electrons only, i.e.,
    $(Y_{1})_{\mu\beta}=(Y_{1})_{\tau\beta}= (Y_{3})_{\mu\beta}=(Y_{3})_{\tau\beta} = 0$.
  \item The scalar leptoquark couples to muons only, i.e.,
    $(Y_{1})_{e\beta}=(Y_{1})_{\tau\beta}= (Y_{3})_{e\beta}=(Y_{3})_{\tau\beta}= 0$.
  \item The scalar leptoquark couples to electrons and muons (no
    taus) with the same rate, i.e.,
    $(Y_{1})_{\mu}=(Y_{1})_{e}$, $(Y_{3})_{\mu\beta}=(Y_{3})_{e\beta}$
    with $(Y_{1})_{\tau\beta}= (Y_{3})_{\tau\beta} = 0$.
\end{enumerate}   
Couplings to the 3rd generation quarks could be sizeable, 
leading to final states involving bottom or even top quarks. 
However, we will limit ourselves in this paper to the study 
of light quarks, i.e. we consider only jets without any flavour 
tags for the quarks.  

We close this section with the short comment that a number of similar
LNV models can easily be constructed, all of which lead in principle
to the same LHC signal.  In fact, the model we have considered in this
section corresponds to a particular example of a $d=9$ short-range
neutrinoless double beta decay operator decomposition that generates
neutrino masses at 2-loop. As can be seen in Table
IV of ref. \cite{Helo:2015fba}, all models in this class have an exotic
coloured fermion that might be pair produced at the LHC.

\section{Montecarlo Simulation} \label{sec:mc}

\begin{figure}[t]
\includegraphics[width=0.44\textwidth]{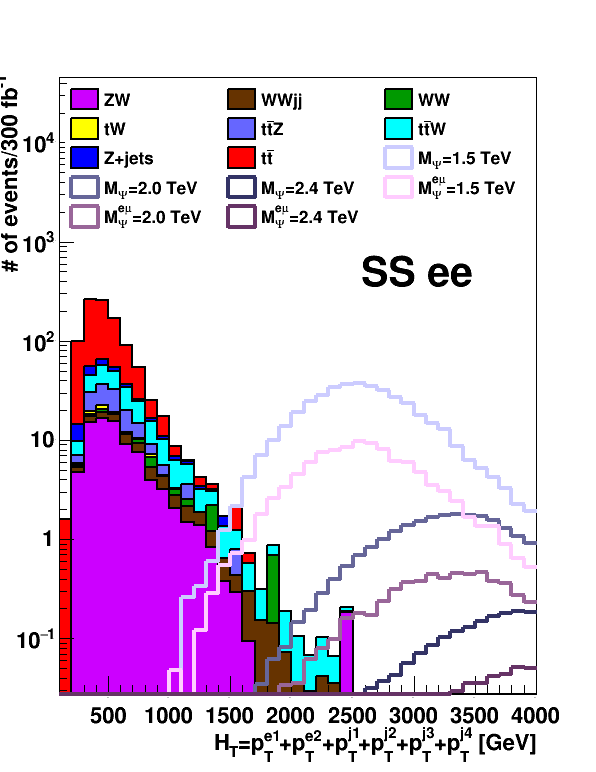}
\includegraphics[width=0.44\textwidth]{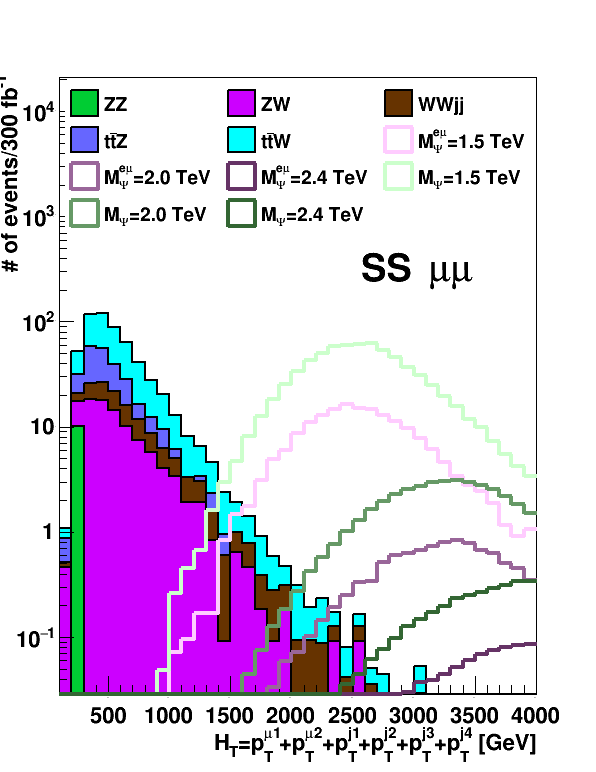}
\captionsetup{justification=raggedright,
singlelinecheck=false
}
\caption{Signal (solid lines) and SM background (stacked histograms)
  $H_T^{ll}$ distributions after the pre-selection cuts described in
  the text are applied. Signal samples for the $ee$ (left), $\mu\mu$
  (right) and $e\mu$ (both plots) scenarios are shown. Only a few
  signal mass points are shown for the sake of
  clarity.} \label{plotee}
\end{figure}

\begin{figure}[h]
\includegraphics[width=0.44\textwidth]{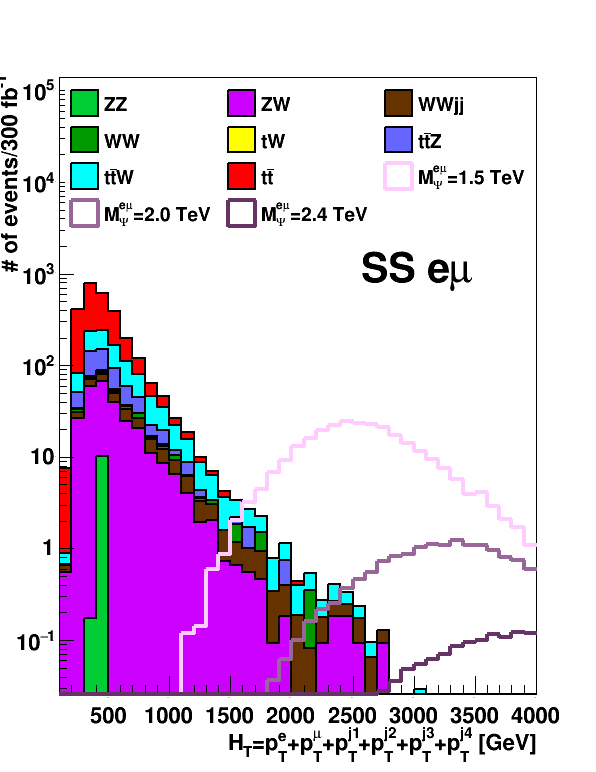}
\captionsetup{justification=raggedright,
singlelinecheck=false
}
  \caption{Signal (solid lines) and SM background (stacked
    histograms) $H_T^{e\mu}$ distributions after the pre-selection
    cuts described in the text are applied. Only a few signal mass
    points are shown for the sake of clarity.} \label{plotemu}
\end{figure}

In order to estimate the sensitivity reach of our model at the LHC
Run-3 (i.e. an integrated luminosity of 300/fb and a center of mass
energy of $\sqrt{s}=13$~TeV), we have performed realistic detector
level simulations of the same-sign (SS) dilepton plus four hard jets
final state signal, shown in figure (\ref{fig:diagLHC}), as well as of the most relevant SM
backgrounds. Our signals correspond to the three benchmark scenarios
described in section \ref{sec:model} with the mass of the colour octet
fermion $m_{\Psi}$ varying in the range [1.5, 2.9] TeV, while the
scalar leptoquark mass is supposed to always be larger than
$m_{\Psi}$, and is therefore off-shell in the $\Psi$ decay. Signal
and SM background events were generated at parton level using
Madgraph5 v2.3.3 \cite{Alwall:2011uj}.

For the simulation of the signal, we have written a private model file
and implemented it in SARAH \cite{Staub:2012pb,Staub:2013tta}. SARAH
then allows to generate automatically a version of SPheno
\cite{Porod:2003um,Porod:2011nf}, with which a numerical calculation
for different observables, such as masses and decay branching ratios
can be done.

For the SM processes, we used the built-in model available in
MadGraph, including the following backgrounds $t\bar{t}, Z,
W^{\pm}W^{\pm}, ZZ, ZW^{\pm},W^{\pm}W^{\mp}, t\bar{t}Z,
t\bar{t}W^{\pm}$ plus 2-4 additional final state partons and $Z$, $W$ and $t$ decaying to leptons ($e,\mu$). Parton
showers and particle decays were generated with Pythia 6.4
\cite{Sjostrand:2000wi,Sjostrand:2006za} and the matrix element to
parton shower matching procedure (MLM) was implemented when generating
SM background processes in order to avoid double counting of the
radiated partons. The interaction of final state particles with the
detector and their reconstruction was simulated using Delphes 3
\cite{Ovyn:2009tx}, configured to replicate the ATLAS detector layout
and its performance. Jets were reconstructed with the anti-$k_t$
algorithm using a cone size of $R=0.4$.  Note that most of the SM
background processes listed above produce opposite-sign (OS) dilepton
final states, while we consider only SS dilepton final states in our
analysis. We must include these processes when we have final states
containing electrons, due to the charge flip effect\footnote{In which
  the electron charge is wrongly tagged, an effect caused by
  electron-positron pair production in hard bremsstrahlung radiation
  emission.}. We expect indeed a significant OS event contamination
into the SS region. In order to account for this effect in final
states containing electrons, we reweighted our events using the charge
flip probability measured by ATLAS (see figure 2a. of
\cite{TheATLAScollaboration:2013jha}), parametrized in electron $p_T$
and $\eta$. We did not consider background contributions from QCD-jets
faking leptons in our analysis, since we expect these to be negligible
for very high-$p_T$ electrons and muons, as those produced in our
signal.

In order to validate our simulated SM backgrounds against measured quantities, we used 
those already performed at the LHC Run 2 by ATLAS and compared the simulated
distribution of number of jets for $t\bar{t}$ and $Z+jets$, with the
measurements performed in \cite{Aaboud:2016pbd} and
\cite{Aaboud:2017hbk}, respectively.  After a simple linear rescaling
of the simulated distributions to the partial Run 2 luminosities, and
applying the selection cuts used in those studies, we compared our
number of jets distributions with the distributions measured by ATLAS
and derived the correction factors needed to account for the observed
differences. The correction was applied to our simulated $Z+jets$ and
$t\bar{t}$ samples as a function of the number of jets, before the
pre-selection used in this analysis. Other samples were not corrected
in this way since the corresponding measurements were not available.
The pre-selection cuts we applied were similar for all final states
considered:
\begin{enumerate}
\item A pair of same-sign (SS) reconstructed leptons with
  $p_T>20$~GeV.
\item In the case of $ee$ final state, we also require the dilepton
  mass to be larger than $110$~GeV, in order to further suppress
  $Z+jets$ background.
\item At least four jets are required, each of them satisfying
  $p_T^{j}>20$~GeV.
\end{enumerate}
As a discriminating variable we used the scalar sum of the hardest
leptons and jets in the event, $H_T\equiv
p_T^{\ell_1}+p_T^{\ell_2}+p_T^{j_1}+p_T^{j_2}+p_T^{j_3}+p_T^{j_4}$. We
found the significance is larger when using $H_T$ compared with
$m_{\ell jj}$ or $m_{\ell \ell jjjj}$ invariant mass distributions,
which are both affected by a large combinatorial
background, due to the many ways the jets can be paired with the leptons. $H_T$ distributions after pre-selection for $ee$, $\mu\mu$
are shown in figure (\ref{plotee}) and $e\mu$ in figure
(\ref{plotemu}).

\section{Results} \label{sec:results}

In this section, we use the results from section~\ref{sec:mc} to
estimate future limits on the $\Psi\rightarrow \ell j j$
($\ell = e,\mu$) branching fraction, for each of the benchmark
scenarios described in section~\ref{sec:model}.  Figures
(\ref{plotee}) and (\ref{plotemu}) show the number
of signal and background events, before the $H_T$ cut, for the $e^\pm
e^\pm j j j j$, $\mu^\pm \mu^\pm j j j j$ and $e^\pm \mu^\pm j j j j$
final states. Depending on the final state, the signals
are shown for the scenarios where the scalar leptoquark couples to
electrons only ($ee$), muons only ($\mu\mu$), or to electrons and
muons with the same strength ($e\mu$).

While the $ee$ ($\mu\mu$) scenario is constrained by the $e^\pm e^\pm j
j j j$ ($\mu^\pm \mu^\pm j j j j$) final state only, the $e\mu$
scenario is constrained by the three final states: $e^\pm e^\pm j j j
j$, $\mu^\pm \mu^\pm j j j j$ and $e^\pm \mu^\pm j j j j$.  For an
off-shell scalar leptoquark ($S_{LQ}$), the number of signal events is
given by
\begin{equation}
S = \sigma(g g \rightarrow \Psi\Psi) \times 
\mbox{Br}(\Psi\rightarrow \ell j j) \times 
\mbox{Br}(\Psi\rightarrow \ell' j j) \times \mathcal{L}_{int}\times \eta,
\end{equation}
where $\ell,\ell'=e,\mu$, $\mathcal{L}_{int} = 300\mbox{ fb}^{-1}$ and
$\eta$ is our cut efficiency. (See the discussion in the previous 
section and the tables in the appendix.)

\begin{figure}[t]
\includegraphics[width=0.48\textwidth]{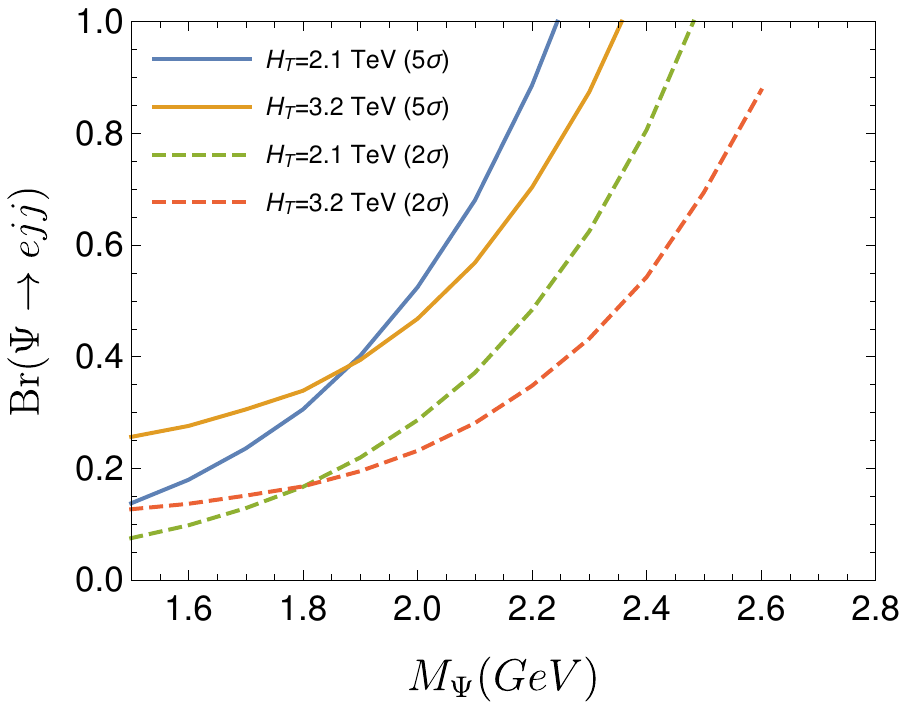}
\includegraphics[width=0.48\textwidth]{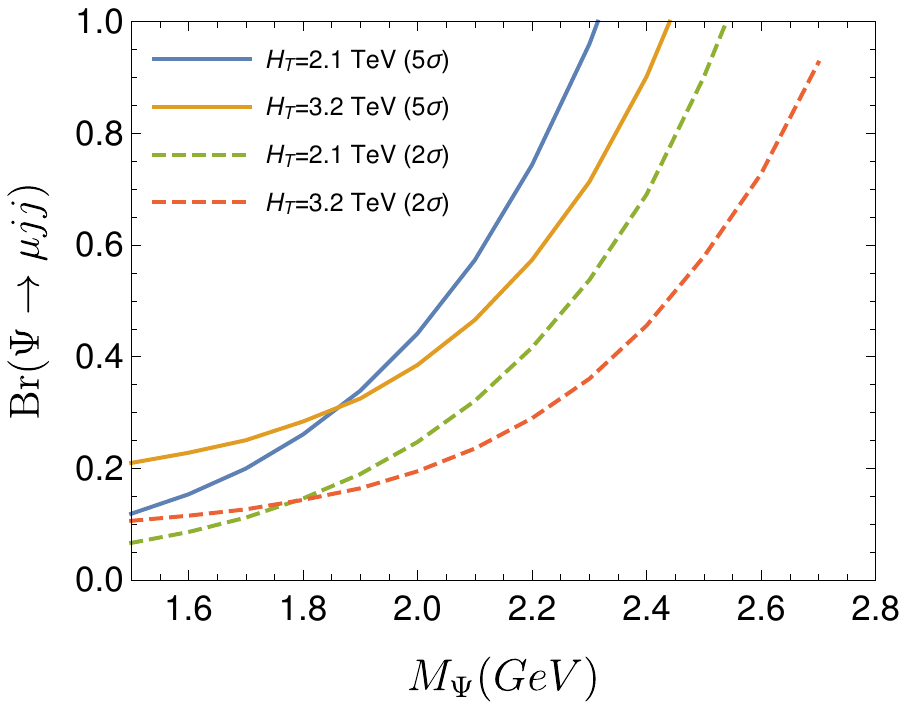}
\captionsetup{justification=raggedright,
singlelinecheck=false
}
\caption{Minimum $\mbox{Br}(\Psi\rightarrow \ell jj)$ for discovery
  (solid lines) and $2\sigma$-limits (dashed lines) as a function of
  $m_\Psi$, for two values of the $H_T$ cut. Left for $\ell=e$ (in the
  scenario where the scalar leptoquark couples to electrons only) and
  right for $\ell=\mu$ (in the scenario where the scalar leptoquark
  couples to muons only). The red dashed line does not reach
  $\mbox{Br}=1$, because we restrict the curves to regions with more
  than one signal event.\label{limitee}}
\end{figure}

\begin{figure}[t]
\includegraphics[width=0.48\textwidth]{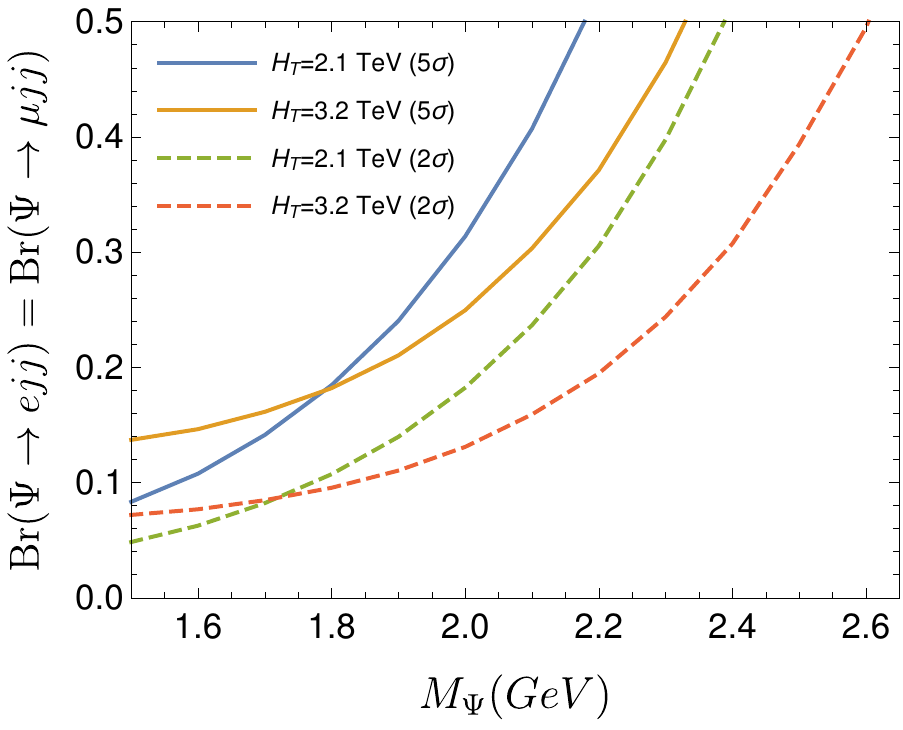}
\captionsetup{justification=raggedright,
singlelinecheck=false
}
\caption{Minimum $\mbox{Br}(\Psi\rightarrow e jj) =
  \mbox{Br}(\Psi\rightarrow \mu jj) $ for discovery (solid lines) and
  $2\sigma$-limits (dashed lines) as a function of $m_\Psi$, for two
  values of the $H_T$ cut, in the scenario where the scalar leptoquark
  couples to electrons and muons with the same
  strength.\label{limitemu}}
\end{figure}

The number of signal events for both the $ee$ and $\mu\mu$ scenarios
were obtained assuming \mbox{$\mbox{Br}(\Psi \rightarrow \ell j j) = 1$},
while for the $e\mu$ scenario we assumed $\mbox{Br}(\Psi \rightarrow e j j)=
\mbox{Br}(\Psi \rightarrow \mu j j) = 1/2$. In order to find the discovery
reach and forecasted limits, we scale these results accordingly and find the
minimum value of the branching ratio required to get a $5\sigma$ or
$2\sigma$ significance.
For the significance, $Z$, we use the expression:
\begin{equation}
 Z = \sqrt{2 \left[(S+B)\times \ln\left(1+\frac{S}{B}\right)-S \right]},
\end{equation}
where $S$ is the number of signal events and $B$ is the number of
background events.
We restrict our analysis to regions with at least one signal event.

Our results are shown in figures (\ref{limitee}) and
(\ref{limitemu}). The three different plots correspond to the three
different scenarios previously discussed: $ee$, $\mu\mu$ and $e\mu$.
In each plot, we show the minimum branching fraction for
$5\sigma$-discovery (solid lines) and $2\sigma$-limits (dashed lines)
as a function of the mass of the colour-octet fermion $\Psi$.  From
these plots one can see that larger values of the $H_T$ cut give
better sensitivities in regions of parameter space with larger values
of $m_\Psi$ (removing most of the backgrounds), while lower values of
$H_T$ are better for smaller values of $m_\Psi$. For this reason, for
each scenario we show our results for two different values of the
$H_T$ cut. As expected, for high (low) masses the expected limits are
stronger (weaker) for lower (higher) values of $H_T$. Depending on the
scenario, the LHC with 300/fb should be able to discover (give limits
for) the colour octet fermion with masses in the range $2.3-2.4$ TeV
($2.6-2.7$ TeV).

Before closing, we want to briefly discuss the LNV searches by CMS
\cite{Sirunyan:2018pom} and ATLAS \cite{Aaboud:2018spl}, cited in the
introduction. Both experiments search for $\ell^{\pm}\ell^{\pm}jj$
final states. Since kinematics and backgrounds are different in this
search relative to the $2\ell 4j$ signal that we are interested in,
limits from these searches can not be straightforwardly converted into
limits on the model considered in this paper. Based on cross sections
alone, from the limits given in figure (5) of \cite{Sirunyan:2018pom}, we
guess-timate that these searches should be able to probe coloured
octet masses roughly up to $m_{\Psi} \sim 2$ TeV. We want to stress,
however, that only a dedicated analysis by the experimental
collaborations can derive the correct limits. Thus, there is ample
room for improving the LHC searches for the LNV model studied in this
paper.

\section{Conclusions\label{sec:conclusions}}

We have studied the potential of the LHC Run-3 to probe lepton number
violation. For our numerical study we have used a particular 2-loop
neutrino mass model. We focussed on a model variant in which the
standard model is extended with two new particles, both singlet under
the SU(2) group: a color-triplet scalar leptoquark $S_{LQ}$ and a
Majorana color-octect fermion $\Psi$. This model is one example of a
model class with a LNV signal at the LHC consisting of same-sign
dileptons plus 4 jets.

We have considered three different benchmark scenarios to take into
account different lepton flavour signals at the LHC: 1) $S_{LQ}$
couples to electrons only, 2) $S_{LQ}$ couples to muons only and 3)
$S_{LQ}$ couples to electrons and muons with the same strength. In view of
the large neutrino mixing angles observed in neutrino oscillation
experiments, one expects benchmark (3) to be the most realistic
one. In order to estimate the sensitivity reach for the LHC, we have
performed realistic detector level simulations for the signal, as well
as for the most relevant SM backgrounds.  We have found that the LHC
should be able to discover the colour octet fermion up to masses in
the range (2.3-2.4) TeV, or derive limits on this model up to masses
order (2.6-2.7) TeV, the exact number depending on the lepton flavour
composition of the final states.

In closing, we would like to point out again that this model is one
example from a large class of models in which the kinematics is
different from the one used by ATLAS \cite{Aaboud:2018spl} and CMS
\cite{Sirunyan:2018pom} searches for LNV in the left-right symmetric
model. Whereas in the left-right model $p_{\ell\ell jj}^2$ should peak at the
mass of the $W_R$, in the model we have discussed the coloured octets are
pair-produced and each decays to $\ell jj$. Thus, there is no ``mass peak''
in the $p_{\ell\ell 4j}^2$ distribution. Instead, we found that the maximum
sensitivity for the LHC can be obtained from studying the
variable $H_{T}$, as discussed in the section \ref{sec:mc}.

\section{Appendix}
\label{sec:app}
In this section we include the expected (simulated) signal and background yields (weighted for 300/fb) for the cut based analysis developed in this work. The tables are organized as follow: Signal and background yields are separated by the horizontal line at mid height, above and below respectively. The first column label the generated processes, the signal points are labelled by the $\Psi$ mass, and the scenario is shown between parenthesis. Numbers are quoted for only a few signal mass points. The second column contain the expected number of events at the target integrated luminosity ($\mathcal{L}_{int}=$300/fb), calculated using the cross section obtained with MadGraph ($\sigma_{MG}$) for each process. The subsequent columns show the remaining event yields after each pre-selection cut is applied (see section \ref{sec:mc}), while the last two columns corresponds to the looser and harder $H_{T}$ cuts, respectively. After column three the event yields also contain the different corrections applied as explained in section \ref{sec:mc}.

\begin{table}[H]
{\scriptsize
\begin{tabular}{| c | c | c | c | c | c | c  | c  | c |} 
\hline 
$m_{\psi}$ TeV (scenario) & $\sigma_{MG}\times \mathcal{L}_{int}$ & ee & $\ell^{\pm}\ell^{\pm}$ & $\geq 4~$ jets& $m_{ee}\!>\!$110~GeV & $H_T\!>$2.1~TeV  & $H_T\!>$3.2~TeV \\ 
\hline 
2.4 ($e\mu$) & 18.05 & 1.60 & 0.80 & 0.79 & 0.78 & 0.78 & 0.68 \\ 
2.0 ($e\mu$) & 146.33 & 13.07 & 6.53 & 6.45 & 6.41 & 6.31 & 3.82 \\ 
1.5 ($e\mu$) & 2395.50 & 218.31 & 109.16 & 107.48 & 106.04 & 90.98 & 13.44 \\ 
2.4 ($ee$) & 18.11 & 6.43 & 3.21 & 3.17 & 3.15 & 3.14 & 2.76 \\ 
2.0 ($ee$) & 146.64 & 51.72 & 25.86 & 25.50 & 25.29 & 24.93 & 15.18 \\ 
1.5 ($ee$) & 2396.69 & 867.22 & 433.61 & 427.69 & 421.70 & 362.93 & 50.62 \\ 
\hline 
$ZZ$ & 3.36$\times 10^6$ & 4.99$\times 10^4$ & 1037.84 & 42.63 & 0.00 & 0.00 & 0.00 \\ 
$ZW$ & 7.29$\times 10^4$ & 9537.70 & 2631.64 & 148.83 & 82.36 & 0.18 & 0.00 \\ 
$WWjj$ & 2059.42 & 151.19 & 151.19 & 26.62 & 18.43 & 0.14 & 0.01 \\ 
$WW$ & 5.58$\times 10^6$ & 6.84$\times 10^4$ & 820.77 & 13.67 & 8.24 & 0.00 & 0.00 \\ 
$tW$ & 1.09$\times 10^6$ & 1.81$\times 10^4$ & 170.38 & 6.22 & 4.71 & 0.00 & 0.00 \\ 
$t\bar{t}Z$ & $1.76\times 10^5$ & 1879.39 & 172.93 & 113.13 & 57.22 & 0.00 & 0.00 \\ 
$t\bar{t}W$ & 4920.41 & 354.99 & 303.27 & 168.71 & 98.96 & 0.22 & 0.02 \\ 
$Z+jets$ & 6.35$\times 10^6$ & 2.76$\times 10^6$ & 2.90$\times 10^4$ & 112.74 & 34.81 & 0.00 & 0.00 \\ 
$t\bar{t}$ & 8.28$\times^6$ & 4.94$\times 10^5$ & 4423.30 & 1204.54 & 709.53 & 0.00 & 0.00 \\ 
\hline 
Total background & 2.49$\times 10^7$ & 3.40$\times 10^6$ & 3.87$\times 10^4$ & 1837.10 & 1014.26 & 0.54 & 0.03 \\ 
\hline 
\end{tabular} 
}
\captionsetup{justification=raggedright,
singlelinecheck=false
}
\caption{Cut flow table showing event yields for signal and SM background processes in the $eejjjj$ final state.} 
\label{table-ee}
\end{table}

\begin{table}[H]
{\scriptsize
\begin{tabular}{| c | c | c | c | c | c  | c  | c  | c  | c  | c |} 
\hline 
$m_{\psi}$ TeV (scenario) & $\sigma_{MG}\times \mathcal{L}_{int}$ & $\mu\mu$ & $\ell^{\pm}\ell^{\pm}$ & $\geq 4~$ jets & $H_T>$2.1~TeV  & $H_T>$3.2~TeV \\ 
\hline 
2.4 ($\mu\mu$) & 18.10 & 11.54 & 5.77 & 5.69 & 5.66 & 4.84 \\ 
2.0 ($\mu\mu$) & 146.62 & 90.94 & 45.47 & 44.94 & 44.19 & 25.97 \\ 
1.5 ($\mu\mu$) & 2399.39 & 1459.49 & 729.75 & 720.87 & 612.38 & 87.69 \\ 
2.4 ($e\mu$) & 18.05 & 2.79 & 1.39 & 1.37 & 1.37 & 1.18 \\ 
2.0 ($e\mu$) & 146.33 & 22.70 & 11.35 & 11.22 & 11.02 & 6.50 \\ 
1.5 ($e\mu$) & 2395.50 & 365.15 & 182.57 & 180.08 & 152.73 & 20.46 \\ 
\hline 
$ZZ$ & 3.36$\times 10^6$ & 8.56$\times 10^4$ & 529.97 & 10.19 & 0.00 & 0.00 \\ 
$ZW$ & 7.29$\times 10^4$ & 1.14$\times 10^4$ & 2157.96 & 95.70 & 0.18 & 0.00 \\ 
$WWjj$ & 2059.42 & 263.06 & 263.06 & 49.24 & 0.42 & 0.03 \\ 
$t\bar{t}Z$ & 1.76$\times 10^5$ & 3052.49 & 192.12 & 115.90 & 0.00 & 0.00 \\ 
$t\bar{t}W$ & 4920.41 & 621.46 & 530.48 & 301.92 & 0.50 & 0.03 \\ 
\hline 
Total background & 3.62$\times 10^6$ & 1.01$\times 10^5$ & 3673.59 & 572.96 & 1.11 & 0.06 \\ 
\hline 
\end{tabular} 
}
\captionsetup{justification=raggedright,
singlelinecheck=false
}
\caption{Cut flow table showing event yields for signal and SM background processes in the $\mu\mu jjjj$ final state.} \label{table-mumu}
\end{table}

\begin{table}[H]
{\scriptsize
\begin{tabular}{| c | c | c | c | c | c  | c  | c  | c  | c  | c  | c |} 
\hline 
$m_{\psi}$ TeV (scenario) & $\sigma_{MG}\times \mathcal{L}_{int}$ & $e\mu$ & $\ell^{\pm}\ell^{\pm}$ & $\geq 4~$ jets & $H_T>$2.1~TeV  & $H_T>$3.2~TeV \\ 
\hline 
2.4 ($e\mu$) & 18.05 & 4.29 & 2.15 & 2.12 & 2.11 & 1.81 \\ 
2.0 ($e\mu$) & 146.33 & 33.72 & 16.86 & 16.69 & 16.43 & 9.86 \\ 
1.5 ($e\mu$) & 2395.50 & 559.55 & 279.78 & 275.97 & 235.27 & 32.98 \\ 
\hline 
$ZZ$ & 3.36$\times 10^6$ & 2293.14 & 855.36 & 10.40 & 0.00 & 0.00 \\ 
$ZW$ & 7.29$\times 10^4$ & 9895.41 & 5017.54 & 264.76 & 0.65 & 0.00 \\ 
$WWjj$ & 2059.42 & 399.49 & 399.49 & 72.20 & 0.58 & 0.04 \\ 
$WW$ & 5.58$\times 10^6$ & 1.79$\times 10^5$ & 1076.52 & 23.70 & 0.27 & 0.00 \\ 
$tW$ & 1.09$\times 10^6$ & 4.61$\times 10^4$ & 246.09 & 11.85 & 0.00 & 0.00 \\ 
$t\bar{t}Z$ & 1.76$\times 10^5$ & 1960.52 & 365.72 & 237.08 & 0.00 & 0.00 \\ 
$t\bar{t}W$ & 4920.41 & 938.40 & 802.70 & 453.05 & 0.69 & 0.05 \\ 
$t\bar{t}$ & 8.28$\times 10^6$ & 1.28$\times 10^6$ & 6065.34 & 1651.22 & 0.00 & 0.00 \\ 
\hline 
Total background & 1.86$\times 10^7$ & 1.52$\times 10^6$ & 1.48$\times 10^4$ & 2724.25 & 2.18 & 0.09 \\ 
\hline 
\end{tabular} 
} 
\captionsetup{justification=raggedright,
singlelinecheck=false
}
 \caption{Cut flow table showing event yields for signal and SM background processes in the $e\mu jjjj$ final state.} \label{table-emu}
\end{table}

\bigskip
\bigskip
\centerline{\bf Acknowledgements}

\medskip
E.C. is supported by chilean grants; FONDECYT No. 11140549, and in part
by CONICYT Basal FB0821.  M.H. is supported by the Spanish grants
SEV-2014-0398 and FPA2017-85216-P (AEI/FEDER, UE) and
PROMETEO/2018/165 (Generalitat Valenciana).  J.C.H. is supported by
Chile grant Fondecyt No. 1161463.  N.N. was supported by FONDECYT
(Chile) grant 3170906 and in part by Conicyt PIA/Basal FB0821.

\bibliographystyle{h-physrev5}

\end{document}